\documentclass[11pt]{iopart}

\usepackage{natbib,url}
\usepackage{graphicx}
\usepackage{iopams}

\begin{document}

\title{Capillary instability  on a hydrophilic stripe}

\author{Raymond L. Speth$^1$ and Eric Lauga$^2$}

\address{$^1$ Department Mechanical Engineering, Massachusetts Institute of Technology, 77 Massachusetts Avenue, Cambridge, MA 02139, USA.}
\vskip 2mm
\address{$^2$ Department of Mechanical and Aerospace Engineering, University of California, San Diego, 9500 Gilman Dr., La Jolla, CA 92093-0411, USA.}
\ead{\mailto{elauga@ucsd.edu}}

\begin{abstract}
A recent experiment showed that cylindrical segments of water filling a hydrophilic stripe on an otherwise hydrophobic surface display a capillary instability when their volume is increased beyond the critical volume at which their apparent contact angle on the surface reaches ninety degrees (Gau {\sl et al.}, {\it Science}, {\bf 283}, 1999).  Surprisingly, the fluid segments did not break up into droplets --- as would be expected for a classical Rayleigh-Plateau instability --- but instead displayed a long-wavelength instability where all excess fluid  gathered in a single bulge along each stripe.  
We consider here the dynamics of the flow instability  associated with this setup.
We perform a linear stability analysis of the capillary flow problem in the inviscid limit. We first confirm previous work showing that that all cylindrical segments are linearly unstable if (and only if) their apparent contact angle is larger than ninety degrees. We then demonstrate that the most unstable wavenumber for the surface perturbation decreases to zero as the apparent contact angle of the fluid on the surface approaches ninety degrees, allowing us to re-interpret the creation of bulges in the experiment  as a zero-wavenumber capillary instability. A variation of the stability calculation is also considered for the case of a hydrophilic stripe located on a wedge-like geometry.

\end{abstract}
\maketitle

\section{Introduction}

Capillary instabilities are phenomena we witness in our daily lives, and their study is a field with a rich history  \cite{eggers97,drazin,degennes_book,pomeauvillermaux}. The classical Rayleigh-Plateau instability refers to the surface-tension induced instability of a cylindrical liquid column. For volume-preserving deformations of sufficiently long wavelengths along a fluid cylinder, the surface area of the fluid can be made to decrease. These deformations lower the surface energy of the fluid and are therefore  favorable, so an infinite cylindrical fluid column is always capillary unstable. For a cylindrical fluid column of radius $R$, density $\rho$ and surface tension $\gamma$, this instability is of  inviscid nature, and  occurs on a typical time scale $\tau_1 \sim (\rho  R^3 /\gamma)^{1/2}$, with the most unstable wavelength being on the order of the column radius  \cite{eggers97,drazin}.

Many variations on this classical result have been considered in the past, and we refer to Ref.~\cite{eggers97} for a review. In the present paper, we consider such instabilities when they occur for a cylindrical segment of fluid  in contact with a solid. A recent experimental investigation on the stability of a cylindrical segment of liquid on a stripe of hydrophilic material on an otherwise hydrophobic surface has shown surprising long-wavelength instabilities. In that case, the fluid segment was observed not  to  break up in many droplets as in the classical Rayleigh-Plateau instability, but instead to form a single  large bulge \cite{gau99}. The purpose of this paper is to study the dynamics of such a surprising capillary instability.

\begin{figure}[b]
\centering
\includegraphics[width=0.6\textwidth]{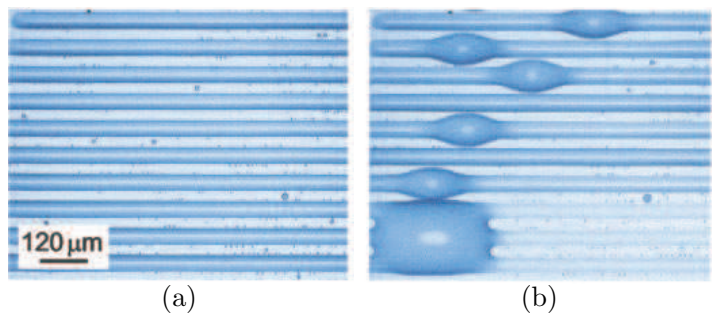}
\caption{The growing instability of a cylindrical segment on a hydrophilic stripe. 
Left: For small enough volume, the cylindrical segments, located along the  hydrophilic stripe and 
separated by hydrophobic stripes, are stable. Right: At a critical value of the fluid volume, corresponding to an apparent contact angle of the fluid of ninety degrees, the cylindrical segment become unstable, and solitary bulges grow. Reprinted from Gau {\sl et al.}, {\it Science}, {\bf 283}, 1999 \cite{gau99}, courtesy of R. Lipowsky and with permission from AAAS [\url{doi:10.1126/science.283.5398.46}].}.
\label{fig:experiment}
\end{figure}

A number of previous studies have approached the problem of flow stability for a cylindrical segment of fluid pinned on a solid substrate.  Davis considered a flow rivulet down an incline, with its contact line pinned along a stripe \cite{davis80}. Using an energy method, it was found that when the interior angle ({\it i.e.} the apparent contact angle of the rivulet on the horizontal surface) was less than $90^\circ$, the rivulet is stable, and it is unstable otherwise. A similar result was later obtained  using energy minimization considerations for a cylindrical interface pinned to a slot \cite{brown80}. In addition, the critical length for the instability of a finite cylinder was found to become infinite at the critical angle of ninety degrees \cite{brown80}. Similar results were later recovered using differential geometry, where volume-preserving perturbations were seen to lead to decrease in surface area
 of the cylindrical filament only in the case where the apparent angle exceeded  $90^\circ$, and for asymptotically large wavelengths near the threshold \cite{roy99,lenz99,lipowsky00,lenz00,brinkmann04}.

\begin{figure}
\centering
\includegraphics[width=1\textwidth]{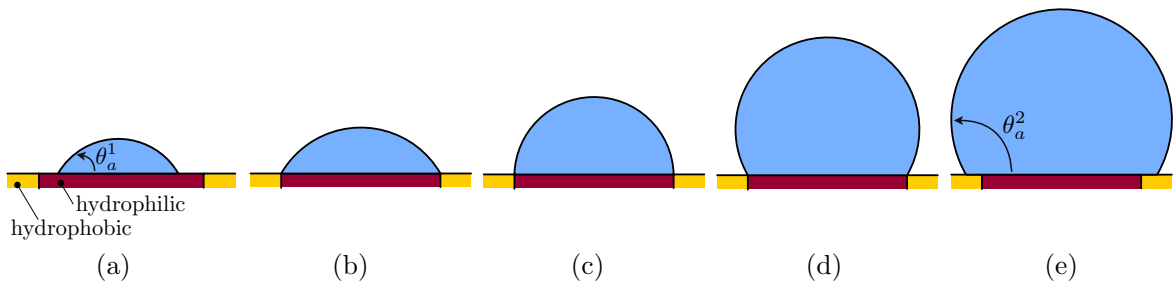}
\caption{Schematic representation of the different stages of  
growth of a cylindrical segment on a hydrophilic stripe in the absence of any capillary instability. 
(a) For small enough volumes, the fluid grows solely on the hydrophilic region, and displays an advancing contact angle,  $\theta_a^1$;
(b) At a critical fluid volume, the  line separating  hydrophilic and hydrophobic domains in reached; 
(c) As the fluid volume continues to increase, the contact line remains pinned, and the apparent angle increases;
(d) The angle stops increasing when the apparent angle reaches the value,  $\theta_a^2$, of the advancing angle on the hydrophobic domain (note that $\theta_a^2>90^\circ>\theta_a^1$); 
(e) As the fluid volume continues to increase, the contact line region now moves into the hydrophobic domains, with angle $\theta_a^2$. In this paper, we show that the cylindrical segment displays capillary instability as soon as (c) is reached (apparent angle of $90^\circ$).}
\label{fig:droplet-growth}
\end{figure}

The experimental observations of  Gau {\sl et al.} \cite{gau99}, which motivate the present study, are reproduced in  Fig.~\ref{fig:experiment}. The fluid (water)  is filling hydrophilic stripes (contact angle $\approx 5^\circ$) on an otherwise hydrophobic substrate  (contact angle $\approx 108^\circ$). The fluid volume is then increased.  In the absence of the observed instability, a schematic representation of the  growth process is represented in Fig.~\ref{fig:droplet-growth}. As the fluid volume  increases, the contact line at the edges of the cylindrical segments of water remains pinned, until the volume of fluid becomes large enough that the apparent contact angle is equal to the advancing contact angle on the hydrophobic surface. At this point, any subsequent volume change would be accompanied by a motion of the contact line into the hydrophobic substrate. The experimental result obtained by Gau {\sl et al.}  is that such a process is unstable. As soon at the apparent contact angle of the fluid segment reaches the critical value of $90^\circ$ (intermediate between the contact angles on both surfaces; see Fig.~\ref{fig:droplet-growth}c), the surface becomes capillary unstable.   This result is consistent with the previous studies discussed above \cite{davis80,brown80,roy99,lenz99,lipowsky00,lenz00}. Surprisingly, and in contrast with a free  fluid cylinder, the  observed unstable mode does not display a wavelength on the order of the cylindrical radius, but instead the wavelength appears to be much larger, and the segment evolves to a state where all the excess fluid gathers on  single bulge (Fig.~\ref{fig:experiment}). Some features of these experiments were later reproduced by Darhuber {\sl et al.}  in their experimental and numerical study of droplet morphologies on chemically patterned surfaces \cite{darhuber00}, and are consistent with the simplified one-dimensional stability study  of Ref.~\cite{schiaffino97}.

The aim of the present work is to focus on the dynamics of the instability process by  performing a linear stability analysis of the cylindrical segment in the experiments of Gau {\sl et al.}, and predicting the dependence of the growth rates and most unstable wavelengths of the unstable modes on  the apparent contact angle of the cylindrical segment. We restrict our study to low viscosity liquids, such as  water, and perform an inviscid study. Physically, inviscid capillary instabilities of a fluid column are expected to occur on a time scale  $\tau_1 \sim (\rho  R^3 /\gamma)^{1/2}$. For comparison, the time scale for viscous effects to propagate diffusively across the width of a fluid column is  $\tau_2\sim R^2/\nu$ where $\nu$ is the kinematic viscosity of the fluid. The inviscid approach will therefore be a reasonable modeling assumption as long as $\tau_1\ll \tau_2$, which is equivalent to $R \gg \ell$, with $\ell =\rho \nu^2/ \sigma$ is the Ohnesorge length scale of the fluid. For water, $\ell$ is on the tens of nanometers, which is much smaller than the typical cross sectional size in the experiments of Gau {\sl et al.} \cite{gau99} (tens of microns; see Fig.~\ref{fig:experiment}). As we show below, within these assumptions, the most unstable wavelength for the capillary instability of the cylindrical segment tends toward infinity as the contact angle approaches $90^\circ$, thereby allowing us to re-interpret the creation of bulges in the experiment  as a zero-wavenumber capillary instability \cite{gau99}.

\section{Setup and linear stability}

\begin{figure}
\centering
\includegraphics[width=0.35\textwidth]{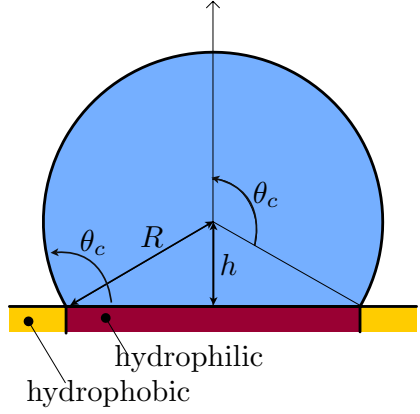}
\caption{Setup for the inviscid stability calculation. We consider a cylindrical segment with a pinned contact line, displaying an apparent contact angle  $\theta_{c}\geq90^\circ$. The radius of the cylindrical segment is denoted $R$, and the height of its center about the surface is denoted $h$.}
\label{fig:droplet-flat}
\end{figure}

The geometrical setup for our linear stability calculation is illustrated in Fig.~\ref{fig:droplet-flat}. 
The basic state is a cylindrical segment of fluid of radius $R$, whose two-dimensional contact line is pinned along a stripe. The apparent contact angle of the fluid at the contact line is denoted $\theta_c$. We assume $\theta_c$ to sufficiently larger (smaller) than the advancing angle on the hydrophilic (hydrophobic) substrate so that we can safely assume that during the initial stages of the instability the contact line remains pinned along the same location. In addition, since we know from earlier work that the case where $\theta_c < 90^\circ$ is stable  \cite{davis80,brown80,roy99,lenz99,lipowsky00,lenz00}, we will focus here on determining the dynamics of the instability in the case where $\theta_c \geq 90^\circ$.

\subsection{Governing equations and linearization}
Assuming the liquid is both incompressible
and inviscid, its motion is prescribed by the Euler equation and the
continuity equation,
\begin{eqnarray}
\rho\left(\frac{\partial\mathbf{u}}{\partial t}+\mathbf{u}\cdot\nabla\mathbf{u}\right)  =  -\nabla p\label{eq:euler},\\
\nabla\cdot\mathbf{u}  =  0,\label{eq:continuity}
\end{eqnarray}
where $\bf u$ and $p$ denote the velocity and pressure fields respectively. Since there is no flow in the basic state, the linearized equations
are simply\begin{eqnarray}
\rho\frac{\partial\mathbf{u}'}{\partial t}  =  -\nabla p'\label{eq:euler-linear},\\
\nabla\cdot\mathbf{u}'  =  0\label{eq:continuity-linear},
\end{eqnarray}
where primes denotes small deviations from the basic state.
We employ cylindrical coordinates, and use the center of the cylindrical segment as the origin of our coordinate system; the variable $z$ denotes therefore the coordinate along the stripe, and $\theta=0$ denotes in the vertical direction. 
Let the radius of the free surface be parameterized as $\xi(t,\theta,z)$.
Since the radius of the free surface is $R$ in the basic state, then
the perturbation in the position of the free surface is $\xi'=\xi-R$.
The free surface moves with the local velocity of the fluid, which after linearization is written as
\begin{equation}
u_{r}'|_{r=R}=\frac{\partial\xi'}{\partial t}\label{eq:bc-surface-kinematic}\cdot
\end{equation}
Additionally, at the contact points, the position of the contact point
is fixed, giving the condition 
\begin{equation}
u'_{r}|_{\{r=R, \theta=\pm \theta_{c}\}}=0.
\end{equation}
 The jump in pressure across the free surface is related to the
curvature of the surface and the surface tension as \begin{equation}
p-p_{\infty}=\gamma\nabla\cdot\mathbf{n},
\end{equation}
where the outward  surface normal is given by 
 \begin{equation}
\mathbf{n}=\frac{1}{\sqrt{1+\left(\frac{\partial\xi}{\partial z}\right)^{2}+\frac{1}{r^{2}}\left(\frac{\partial\xi}{\partial\theta}\right)^{2}}}\left(\mathbf{e}_{r}-\frac{1}{r}\frac{\partial\xi}{\partial\theta}\mathbf{e}_{\theta}-\frac{\partial\xi}{\partial z}\mathbf{e}_{z}\right),\end{equation}
and the divergence of the normal is \begin{equation}
\nabla\cdot\mathbf{n}=\frac{\partial n_{z}}{\partial z}+\frac{1}{r}\frac{\partial}{\partial r}\left(rn_{r}\right)+\frac{1}{r}\frac{\partial n_{\theta}}{\partial\theta}\cdot
\end{equation}
When linearized, this boundary condition becomes\begin{equation}
p'|_{r=R}=-\gamma\left(\frac{\xi'}{R^{2}}+\frac{\partial^{2}\xi'}{\partial z^{2}}+\frac{1}{R^{2}}\frac{\partial^{2}\xi'}{\partial\theta^{2}}\right)\label{eq:bc-surface-stress}\cdot
\end{equation}
At the solid surface, the normal component of the velocity vanishes, and therefore
 \begin{equation}
\mathbf{u}'\cdot\mathbf{n}=0.\label{eq:bc-wall-velocity}
\end{equation}

\subsection{Normal modes}

Taking the divergence of the linearized Euler equation, Eq.~\ref{eq:euler-linear},
gives the Laplace equation for the pressure, 
\begin{equation}
\nabla^{2}p'=0.
\label{eq:laplace}
\end{equation}
Considering normal modes for $p'$, $\mathbf{u}'$ and $\xi'$ such
that \begin{eqnarray}
p'(r,\theta,z,t) & = & \hat{p}(r,\theta)e^{st+ikz},\\
\mathbf{u}'(r,\theta,z,t) & = & \hat{\mathbf{u}}(r,\theta)e^{st+ikz},\\
\xi'(\theta,z,t) & = & \hat{\xi}(\theta)e^{st+ikz},\end{eqnarray}
Eq.~\ref{eq:laplace} becomes\begin{equation}
\frac{\partial^{2}\hat{p}}{\partial r^{2}}+\frac{1}{r}\frac{\partial\hat{p}}{\partial r}+\frac{1}{r^{2}}\frac{\partial^{2}\hat{p}}{\partial\theta^{2}}-k^{2}\hat{p}=0\label{eq:helmholtz}.\end{equation}
Substituting these normal modes into Eq.~\ref{eq:euler-linear}
relates the pressure to the velocity,\begin{equation}
\rho s\hat{\mathbf{u}}=-\frac{\partial\hat{p}}{\partial r}\mathbf{e}_{r}-\frac{1}{r}\frac{\partial\hat{p}}{\partial\theta}\mathbf{e}_{\theta}-ik\hat{p}\mathbf{e}_{z}\label{eq:euler2},\end{equation}
allowing the boundary conditions to be expressed solely in terms of
$\hat{p}$. The boundary condition for the motion of the free surface,
Eq.~\ref{eq:bc-surface-kinematic}, then becomes 
\begin{equation}
-\frac{1}{\rho s}\frac{\partial\hat{p}}{\partial r}=s\hat{\xi}.
\label{eq:bc-surface-kinematic2}
\end{equation}
The stress boundary condition on the free surface, Eq.~\ref{eq:bc-surface-stress}, 
becomes
\begin{equation}
\hat{p}=-\gamma\left(\frac{1}{R^{2}}-k^{2}+\frac{1}{R^{2}}\frac{\partial^{2}}{\partial\theta^{2}}\right)\hat{\xi}\label{eq:bc-surface-stress2}.\end{equation}
Combining Eqs.~\ref{eq:bc-surface-kinematic2} and \ref{eq:bc-surface-stress2}
to eliminate $\hat{\xi}$ leads to \begin{equation}
s^{2}\hat{p}=\frac{\gamma}{\rho R^{2}}\left(1-k^{2}R^{2}+\frac{\partial^{2}}{\partial\theta^{2}}\right)\frac{\partial\hat{p}}{\partial r}\label{eq:bc-pressure-free}\cdot
\end{equation}
Next, we need to express the no-penetration boundary condition at
the wall in terms of $\hat{p}$.  Using the relations between $\hat{p}$ and $\mathbf{\hat{u}}$ from Eq.~\ref{eq:euler2},
the no-penetration condition, Eq.~\ref{eq:bc-wall-velocity}, 
may be written as \begin{equation}
\frac{\partial\hat{p}}{\partial r}n_{r}+\frac{1}{r}\frac{\partial\hat{p}}{\partial\theta}n_{\theta}=0\label{eq:bc-wall-velocity2},
\end{equation}
along the solid substrate. 
Finally, the condition that the contact points of the free surface
are stationary requires \begin{equation}
\frac{\partial\hat{p}}{\partial r}=0\label{eq:xi-stationary},
\end{equation}
at the contact points. The complete eigenvalue problem to solve for the pressure field is therefore given by Eq.~\ref{eq:helmholtz}, together with the boundary conditions provided by Eqs.~\ref{eq:bc-pressure-free},
\ref{eq:bc-wall-velocity2} and \ref{eq:xi-stationary}.

We may solve Eq.~\ref{eq:helmholtz}
by the method of  separation of variables, letting $\hat{p}(r,\theta)=F(r)G(\theta)$.
Applying this separation gives\begin{equation}\label{eq:bessel}
\frac{F''}{F}+\frac{F'}{rF}-\frac{k^{2}}{r^{2}R}=-\frac{G''}{G}=\lambda^{2}.\end{equation}
We see that Eq.~\ref{eq:bessel} for $F$ is the modified Bessel differential equation,
whose solutions are modified Bessel functions of the first and second
kind, \begin{equation}
F(r)=C_{1}I_{\lambda}(kr)+C_{2}K_{\lambda}(kr)\label{eq:bessel-soln},\end{equation}
whereas the general solution to Eq.~\ref{eq:bessel} for $G$ is \begin{equation}
G(\theta)=C_{3}\sin\lambda\theta+C_{4}\cos\lambda\theta\label{eq:harmonic-soln}.
\end{equation}
Applying the periodicity condition, that is, $G(\theta)=G(\theta+2\pi)$
to Eq.~\ref{eq:harmonic-soln} gives that $\lambda_{n}=n$ where
$n=0,1,2,...$. Applying the condition that $\hat{p}$ is finite at
$r=0$ to Eq.~\ref{eq:bessel-soln} eliminates $K_{\lambda}$
as a solution for $F$. A general solution for $\hat{p}$ may then
be written as \begin{equation}
\hat{p}\left(r,\theta\right)=\sum_{n=0}^{\infty}I_{n}(kr)\left(A_{n}\sin n\theta+B_{n}\cos n\theta\right),
\end{equation}
where $A_{n}$ and $B_{n}$ are series of unknown constants to be
determined by the boundary conditions. Applying the boundary condition
at the free surface, Eq.~\ref{eq:bc-pressure-free}, with $r=R$
and $\left|\theta\right|<\theta_{c}$ produces
\begin{eqnarray}\label{eq:bc-eig1}
s^{2}\sum_{n=0}^{\infty}I_{n}(kR)\left(A_{n}\sin n\theta+B_{n}\cos n\theta\right)\\ =\frac{\gamma}{\rho R^{2}}\sum_{n=0}^{\infty}k\left(1-k^{2}R{}^{2}-n^{2}\right)I_{n}'(kR)\left(A_{n}\sin n\theta+B_{n}\cos n\theta\right).\nonumber
\end{eqnarray}
The no-penetration condition at the lower wall, Eq.~\ref{eq:bc-wall-velocity2},  is given by
\begin{equation}
\frac{\partial\hat{p}}{\partial r}+\frac{\sin\theta}{h}\frac{\partial\hat{p}}{\partial\theta}=0\label{eq:no-penetration},\end{equation}
where $h$ is the vertical separation between the origin and the wall.  Applying this condition using $r=-h/\cos\theta$ and $\theta_{c}<\left|\theta\right|<\pi$
produces
\begin{eqnarray}\label{31}
\sum_{n=0}^{\infty}kI_{n}'\!\left(-\frac{kh}{\cos\theta}\right)\left(A_{n}\sin n\theta-B_{n}\cos n\theta\right)\\ =- \sum_{n=0}^\infty\frac{n\sin\theta}{h}I_{n}\!\left(-\frac{kh}{\cos\theta}\right)\left(A_{n}\cos n\theta+B_{n}\sin n\theta\right)\label{eq:bc-eig2}.
\nonumber
\end{eqnarray}
At the contact points, $r=R$, $\left|\theta\right|=\theta_{c}$,
Eq.~\ref{eq:xi-stationary} becomes\begin{equation}
\sum_{n=0}^{\infty}kI_{n}'(kR)\left(A_{n}\sin n\theta+B_{n}\cos n\theta\right)=0\label{eq:bc-eig3}.
\end{equation}

\subsection{Eigenvalue problem}

Together, Eqs.~\ref{eq:bc-eig1}, \ref{31} and \ref{eq:bc-eig3}
represent a generalized eigenvalue problem, where the eigenvalue is the square of
the growth rate, $s^{2}$, and the eigenvector is composed of the
series of constants $A_{n}$ and $B_{n}$. Unlike many other eigenvalue
problems which arise in separation of variables solutions, these equations
cannot be solved in the usual manner, that is by multiplying by one
of the modes, integrating over the domain of $\theta$ and applying
the orthogonality of the modes to generate analytical expressions
for the constants $A_{n}$ and $B_{n}$. This is because the boundary conditions in the physical problem are of mixed type.

An approximate numerical solution to the eigenvalue problem may be sought by
truncating the series at $n=N$, and evaluating the boundary conditions
at some number of discrete $\theta=\left\{ \theta_{1},\ldots,\theta_{2N+1}\right\} $
to produce a linear system of $2N+1$ equations in $2N+1$ unknowns
($A_{0}$ is irrelevant). The set of $\theta$ are picked to specifically
include $\pm\theta_{c}$ to ensure that the boundary conditions at
those points are met. Formally, the eigenvalue problem of Eqs.~\ref{eq:bc-eig1}---\ref{eq:bc-eig3} may be rewritten as\begin{equation}
\sum_{n=0}^{\infty}A_{n}a_{n}(\theta)+B_{n}b_{n}(\theta)=s^{2}\sum_{n=0}^{\infty}A_{n}\alpha_{n}(\theta)+B_{n}\beta_{n}(\theta),
\end{equation}
where the functions $a_{n}$, $b_{n}$, $\alpha_{n}$ and $\beta_{n}$
are given by
\begin{equation}
a_{n}=\left\{
\begin{array}{ll}
\frac{\gamma}{\rho R^{2}}I_{n}'(kR)k\left(1-k^{2}R^{2}-n^{2}\right)\sin n\theta & \left|\theta\right|<\theta_{c},\\
kI_{n}'\!\left(-\frac{kh}{\cos\theta}\right)\sin n\theta+\frac{n\sin\theta}{h}I_{n}\!\left(-\frac{kh}{\cos\theta}\right)\cos n\theta & \left|\theta\right|=\theta_{c},\\
kI_{n}'(kR)\sin n\theta & \theta_{c}<\left|\theta\right|<\pi,
\end{array}
\right.
\end{equation}
\begin{equation}
b_{n}=
\left\{\begin{array}{ll}
\frac{\gamma}{\rho R^{2}}I_{n}'(kR)\left(1-k^{2}R^{2}-n^{2}\right)\cos n\theta & \left|\theta\right|<\theta_{c},\\
kI_{n}'\!\left(-\frac{kh}{\cos\theta}\right)\cos n\theta-\frac{n\sin\theta}{h}I_{n}\!\left(-\frac{kh}{\cos\theta}\right)\sin n\theta & \left|\theta\right|=\theta_{c},\\
kI_{n}'(kR)\cos n\theta & \theta_{c}<\left|\theta\right|<\pi
,
\end{array}\right.
\end{equation}
\begin{equation}
\alpha_{n}=
\left\{\begin{array}{ll}
I_{n}(kR)\sin n\theta & \left|\theta\right|<\theta_{c},\\
0 & \left|\theta\right|=\theta_{c},\\
0 & \theta_{c}<\left|\theta\right|<\pi,
\end{array}\right.
\end{equation}
and
\begin{equation}
\beta_{n}=
\left\{\begin{array}{ll}
I_{n}(kR)\cos n\theta & \left|\theta\right|<\theta_{c},\\
0 & \left|\theta\right|=\theta_{c},\\
0 & \theta_{c}<\left|\theta\right|<\pi.
\end{array}\right.
\end{equation}
The approximate eigenvalue problem may then be written in matrix form
as 
\begin{eqnarray}
\left[\begin{array}{cccccc}
a_{1}(\theta_{1}) & \cdots & a_{N}(\theta_{1}) & b_{0}(\theta_{1}) & \cdots & b_{N}(\theta_{1})\\
\\\vdots &  & \vdots & \vdots &  & \vdots\\
\\a_{1}(\theta_{2N+1}) & \cdots & a_{N}(\theta_{2N+1}) & b_{0}(\theta_{2N+1}) & \cdots & b_{N}(\theta_{2N+1})\end{array}\right]\left[\begin{array}{c}
A_{1}\\
\vdots\\
A_{N}\\
B_{0}\\
\vdots\\
B_{N}\end{array}\right]\\
=s^{2}\left[\begin{array}{cccccc}
\alpha_{1}(\theta_{1}) & \cdots & \alpha_{N}(\theta_{1}) & \beta_{0}(\theta_{1}) & \cdots & \beta_{N}(\theta_{1})\\
\\\vdots &  & \vdots & \vdots &  & \vdots\\
\\\alpha_{1}(\theta_{2N+1}) & \cdots & \alpha_{N}(\theta_{2N+1}) & \beta_{0}(\theta_{2N+1}) & \cdots & \beta_{N}(\theta_{2N+1})\end{array}\right]\left[\begin{array}{c}
A_{1}\\
\vdots\\
A_{N}\\
B_{0}\\
\vdots\\
B_{N}\end{array}\right]\label{eq:gen-eigenvalue}\nonumber.
\end{eqnarray}
The solutions to this generalized eigenvalue problem may be found
using Matlab or comparable software.

\section{\label{sec:Results}Stability results}

\begin{figure}[t]
\centering
\includegraphics[width=0.5\textwidth]{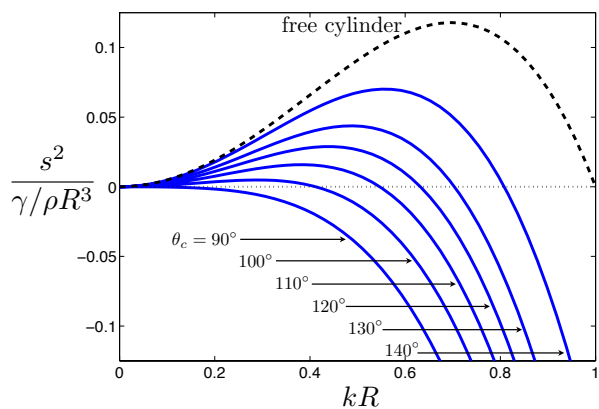}
\caption{Capillary instability of cylindrical segment on a hydrophilic stripe: Square of the dimensionless growth rate, $s^2/(\gamma/\rho R^3)$, as a function of the dimensionless wavenumber, $kR$, for various value of the apparent angle of the stripe, $\theta_c=90$$^\circ$, 100$^\circ$, 110$^\circ$, 120$^\circ$, 130$^\circ$ and 140$^\circ$. We also reproduce as a  dashed line the growth rate for a free cylinder (classical Rayleigh-Plateau instability).}
\label{fig:Growth-rate-segment}
\end{figure}

\begin{figure}[t]
\centering
\includegraphics[width=0.4\textwidth]{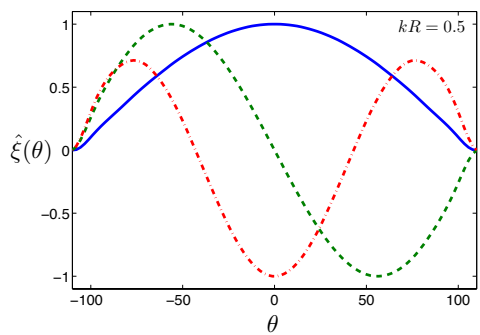}
\caption{
Normalized mode shapes, $\hat\xi(\theta)$, as a function of the angle $\theta$ about the vertical direction, for a cylindrical segment with $\theta_{c}=110^{\circ}$, and wavenumber $kR=0.5$. The three most unstable modes are plotted:  
$s^2/(\gamma/\rho R^3)=0.008$ (solid line, blue; this is the only unstable mode), -2.539 (dashed line, green), -10.11 (dash-dotted line, red).}
\label{fig:modes-segment}
\end{figure}

\begin{figure}[b]
\centering
\includegraphics[width=0.5\textwidth]{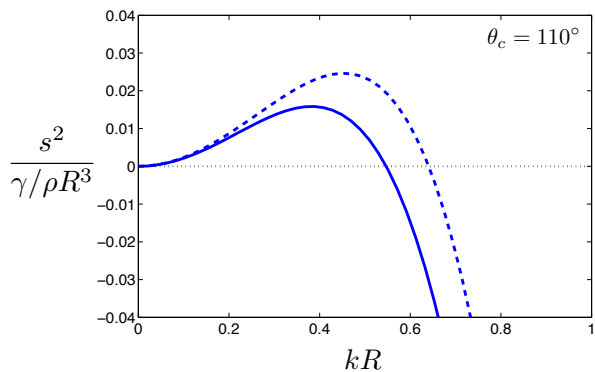}
\caption{Comparison between the inviscid stability calculation from the present paper (solid line) and the one-dimensional inviscid model from Schiaffino {\sl et al.} \cite{schiaffino97} (dashed line) for $\theta_c=110^\circ$.  Square of the dimensionless growth rate of the most unstable mode, $s^2/(\gamma/\rho R^3)$, as a function of the dimensionless wavenumber, $kR$.}
\label{fig:sonin}
\end{figure}

The results of our linear stability calculation are illustrated in Fig.~\ref{fig:Growth-rate-segment}, where we plot 
the square of the dimensionless growth rate of the most unstable mode, $s^2/(\gamma/\rho R^3)$, as a function of the dimensionless  wavenumber of the perturbation, $kR$, for different values of the apparent contact angle $\theta_c$ (solid lines). We also plot for comparison the result for a free cylinder, {\it i.e.}
\begin{equation}
s^{2}=\frac{\gamma}{\rho R^{3}}\frac{I_{0}'(kR)}{I_{0}(kR)}\left(1-k^{2}R^{2}\right) kR,
\end{equation}
which is the classical Rayleigh-Plateau result (dashed line) \cite{eggers97,drazin}. An instability is possible only if there exists a value of $k$ for which a mode of deformation satisfies $s^2 >0$. We see in Fig.~\ref{fig:Growth-rate-segment} that, in accordance with previous work, the cylindrical segments with $\theta_c > 90^\circ$ are always linearly unstable \cite{davis80,roy99,lenz99,lipowsky00,lenz00,brinkmann04}. 
The main result of the paper, as seen in Fig.~\ref{fig:Growth-rate-segment}, is the explicit calculation of the range of unstable wavenumbers (together with the associated growth rates) and in particular the result that this range reaches zero for $\theta_c = 90^\circ$, which coincides with the limit of the stability domain. In other words, in the experiment of Ref.~\cite{gau99}, as soon as the apparent contact angle reaches the critical value of $\theta_c = 90^\circ$ from below, the fluid becomes capillary unstable, but at a wavenumber that is close to zero, corresponding therefore to deformations with infinitely long wavelengths. Consequently, the experimentally observed bulges are the manifestation of a zero-wavenumber capillary instability \cite{gau99}. We further plot in Fig.~\ref{fig:modes-segment}  the shape of the three most unstable modes
for the position of the free surface $\hat{\xi}(\theta)$ for $\theta_{c}=110^\circ$ and $kR=0.5$. Additional modes reflect higher spatial harmonics, all with negative values for $s^{2}$. In no case does
the value of $s^{2}$ become positive for any mode except the first, as is the case for the Rayleigh-Plateau problem \cite{eggers97,drazin}. Finally, we compare in Fig.~\ref{fig:sonin} the results of our stability calculation (solid line) with that of the one-dimensional model of  Ref.~\cite{schiaffino97}  (dashed line) for $\theta_c=110^\circ$. We see that the one-dimensional approximation, although qualitatively similar to the the results of the full calculation,  over-estimates both the growth rate and the most unstable wavenumber.

\section{Cylindrical segment on a wedge}

\begin{figure}[t]
\centering
\includegraphics[width=0.6\textwidth]{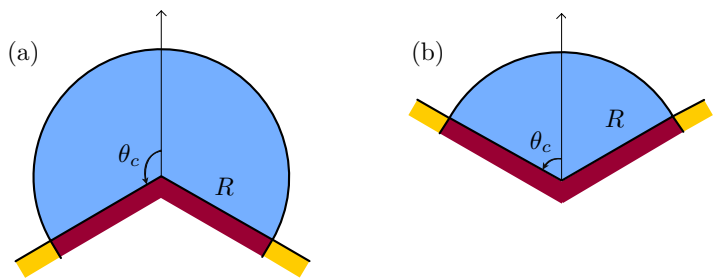}
\caption{Fluid segment on a hydrophilic stripe on a wedge of opening angle $2\theta_c$: (a) $\theta_{c}>90^\circ$
and (b) $\theta_{c}<90^\circ$.}
\label{fig:droplet-wedge}
\end{figure}

Inspired by the results above, we consider now a different geometrical setup, as illustrated in Fig.~\ref{fig:droplet-wedge}. So far we have assumed the substrate to be horizontal. In that case,  when the fluid volume is increased, both  the apparent contact angle of the cylindrical segment on the surface and the cross-sectional shape of the fluid change. We now consider a setup where the contact angle is fixed, while the shape is allowed to change. Specifically, we consider a cylindrical segment of fluid on a hydrophilic stripe in a wedge-like geometry. The volume of the fluid and the opening angle of the wedge, $2\theta_c$, are supposed to be such that the apparent contact  angle of the fluid on the surface always remains to be $90^\circ$. As a result, the center of the cross section of the fluid segment is a circular wedge centered at the tip of the hydrophilic wedge (see Fig.~\ref{fig:droplet-wedge}). This geometry is reminiscent of previous work on droplets in  angular geometries, in both static \cite{langbein90,brinkmann04_EPJE} and flowing  \cite{yang07}  conditions.

We consider therefore the capillary stability of the configuration illustrated in Fig.~\ref{fig:droplet-wedge} with  notation similar to the previous section. We now use a cylindrical coordinate system  with the origin at the point of the wedge, and with $\theta=0$
aligned with one segment of the solid wall and $\theta=2\theta_{c}$ with the other.
The boundary conditions in this geometry may be applied as follows.
Applying the no-penetration boundary condition, Eq.~\ref{eq:bc-wall-velocity2},
to the harmonic equation for the $\theta$-dependence of the solution,
Eq.~\ref{eq:harmonic-soln}, Gives that $C_{3}=0$ and $\lambda_{n}=n\pi/2\theta_{c}$
where $n=0,1,2,\ldots$. As before, requiring that $\hat{p}$ be finite
at $r=0$ requires that $C_{2}=0$. The form of the solution for $\hat{p}$
is then\begin{equation}
\hat{p}(r,\theta)=\sum_{n=0}^{\infty}A_{n}I_{\lambda_{n}}(kr)\cos\frac{n\pi\theta}{2\theta_{c}}\cdot
\end{equation}
The boundary condition along the free surface, \{$r=R$, $0<\theta<2\theta_{c}$\}, 
as given by Eq.~\ref{eq:bc-pressure-free} becomes
\begin{eqnarray}
s^{2}\sum_{n=0}^{\infty}A_{n}I_{\lambda_{n}}(kR)\cos\frac{n\pi\theta}{2\theta_{c}}\\ =\sum_{n=0}^{\infty}\frac{\gamma}{\rho R^{2}}k\left(1-k^{2}R^{2}-\frac{n^{2}\pi^{2}}{4\theta_{c}^{2}}\right)A_{n}I_{\lambda_{n}}'(kR)\cos\frac{n\pi\theta}{2\theta_{c}}\cdot
\label{eq:bc2-eig1}\nonumber
\end{eqnarray}
At the contact points, \{$r=R$, $\theta=0,2\theta_{c}$\}, Eq.~\ref{eq:xi-stationary}
becomes\begin{equation}
\sum_{n=0}^{\infty}A_{n}I_{\lambda_{n}}'(kR)=0\label{eq:bc2-eig2}.
\end{equation}
As above, we obtain a generalized eigenvalue problem for the
eigenvalue $s^{2}$ and the eigenvector $A_{n}$. 
We can combine Eqs.~\ref{eq:bc2-eig1} and \ref{eq:bc2-eig2}
by writing them as\begin{equation}
\sum_{n=0}^{\infty}A_{n}a_{n}(\theta)=s^{2}\sum_{n=0}^{\infty}A_{n}\alpha_{n}(\theta),
\label{eq:eig2-cont}\end{equation}
where the coefficients $a_{n}$ and $\alpha_{n}$ are defined as
\begin{equation}
a_{n}(\theta)=
\left\{\begin{array}{ll}
\frac{\gamma}{\rho R^{2}}k\left(1-k^{2}R^{2}-\frac{n^{2}\pi^{2}}{4\theta_{c}^{2}}\right)I_{\lambda_{n}}'(kR)\cos\frac{n\pi\theta}{2\theta_{c}} & 0<\theta<2\theta_{c},\\
I_{\lambda_{n}}'(kR) & \theta=0\mbox{ or }\theta=2\theta_{c},
\end{array}\right.
\end{equation}
\begin{equation}
\alpha_{n}(\theta)=
\left\{\begin{array}{ll}
I_{\lambda_{n}}(kR)\cos\frac{n\pi\theta}{2\theta_{c}} & 0<\theta<2\theta_{c},\\
0 & \theta=0\mbox{ or }\theta=2\theta_{c}.
\end{array}\right.
\end{equation}
\begin{figure}[t]
\centering
\includegraphics[width=0.6\textwidth]{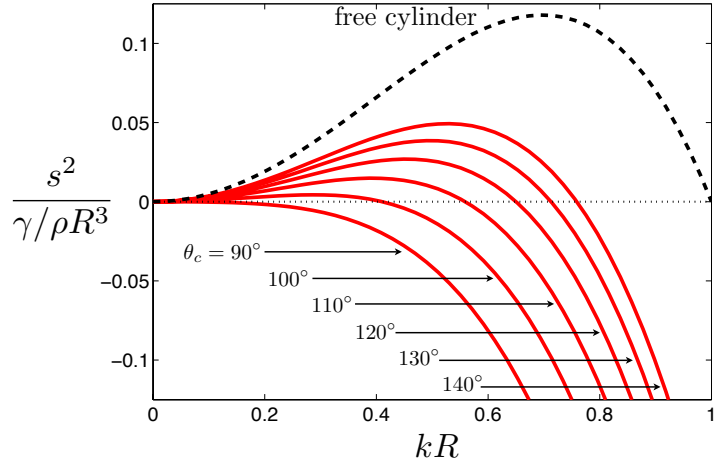}
\caption{Capillary instability of cylindrical segment on a wedge:  Square of the dimensionless growth rate of the most unstable mode, $s^2/(\gamma/\rho R^3)$, as a function of the dimensionless wavenumber, $kR$, for various value of the apparent angle of the stripe, $\theta_c=90$$^\circ$, 100$^\circ$, 110$^\circ$, 120$^\circ$, 130$^\circ$ and 140$^\circ$. We also reproduce in dashed line the growth rate for a free cylinder (classical Rayleigh-Plateau instability).}
\label{fig:Growth-wedge}
\end{figure}
As above, an approximate solution can be found by truncating the series
at $n=N$ and evaluating Eq.~\ref{eq:eig2-cont} at $N+1$ values
of $\theta$ which include $\theta=0$. In matrix form, the truncated
eigenvalue problem is now given by
\begin{eqnarray}
\left[\begin{array}{ccc}
a_{0}(\theta_{1}) & \cdots & a_{N}(\theta_{1})\\
\vdots &  & \vdots\\
a_{0}(\theta_{N+1}) & \cdots & a_{N}(\theta_{N+1})\end{array}\right]\left[\begin{array}{c}
A_{0}\\
\vdots\\
A_{N}\end{array}\right]=\\
s^{2}\left[\begin{array}{ccc}
\alpha_{0}(\theta_{1}) & \cdots & \alpha_{N}(\theta_{1})\\
\vdots &  & \vdots\\
\alpha_{0}(\theta_{N+1}) & \cdots & \alpha_{N}(\theta_{N+1})\end{array}\right]\left[\begin{array}{c}
A_{0}\\
\vdots\\
\nonumber A_{N}\end{array}\right]\cdot\label{eq:gen-eigenvalue2}\end{eqnarray}

By numerically solving this eigenvalue system, we find that such a  cylindrical wedge of fluid is stable as long as the opening of the wedge satisfies $\theta_c<90^\circ$. When $\theta_c\geq 90^\circ$ we find unstable modes, and the square of their dimensionless growth rates are displayed in Fig.~\ref{fig:Growth-wedge} as a function of the dimensionless wavenumber of the perturbation (solid lines). As above, we have included the classical Rayleigh-Plateau result (dashed line). We see that the growth rate of the most unstable modes in this setup (wedge-like stripe) are very similar to the ones obtained in the previous section (horizontal stripe). In particular, we recover the result that in the limit where the angle $\theta_c\to 90^\circ$, the most unstable perturbation wavelength increases to infinity.

\section{Conclusion}

In this paper, we have studied the dynamics of the capillary instability discovered  experimentally  by Gau {\sl et al.} \cite{gau99}. In this work, it was shown that a circular segment of fluid located on a hydrophilic stripe on an otherwise hydrophobic substrate becomes unstable when its volume reaches that at which its apparent contact angle on the surface is  ninety degrees. Instead of   breaking up into droplets, the instability lead to the excess fluid collecting into a single bulge along each stripe. By performing a  linear stability analysis of the capillary flow problem in the inviscid limit, we have first reproduced previous results showing  
that the cylindrical segment are linearly unstable if (and only if) their apparent contact angle is larger than ninety degrees. We have then calculated the growth rate of the instability as a function of the wavenumber of the perturbation, and  shown that  the most unstable wavenumber for the instability --- the one which would therefore be observed in an experimental setting ---  decreases to zero when the apparent fluid contact angle  reaches ninety degrees, allowing us to re-interpret the creation of bulges in the experiment  as a zero-wavenumber capillary instability \cite{gau99}. A variation of the stability calculation was also considered in the case of a hydrophilic stripe located on a wedge-like geometry. 
Since droplets of any shape can now be created experimentally using chemical substrate modification \cite{jokinen08}, the stability of more complex fluid topologies could be analyzed using a framework similar to the one developed here.

\section*{Acknowledgments}
The authors thank Prof. Lipowsky for allowing us to reproduce in Fig.~\ref{fig:experiment} the picture from Ref.~\cite{gau99}. This work was funded in part by the National Science Foundation (grants CTS-0624830 and CBET-0746285 to Eric Lauga).

\bibliographystyle{unsrt}
\bibliography{stripe}

\end{document}